\documentclass[pra,aps,amsmath,amssymb,showkeys]{revtex4}
\usepackage{graphicx}
\newcommand{\hh}{{\mathcal{H}}}

\newcommand{\lnp}{\mathcal{L}}
\newcommand{\lsp}{\mathcal{L}_{+}}
\newcommand{\pen}{\openone}
\newcommand{\bro}{\boldsymbol{\rho}}

\newcommand{\bdl}{\boldsymbol{\delta}}
\newcommand{\bmg}{\boldsymbol{\omega}}
\newcommand{\wbro}{\widetilde{\boldsymbol{\rho}}}
\newcommand{\clb}{\mathcal{B}}
\newcommand{\cli}{{\mathcal{I}}}
\newcommand{\me}{{\mathsf{E}}}

\newcommand{\dn}{{\mathbb{D}}}

\newcommand{\cald}{{\mathcal{D}}}
\newcommand{\calp}{{\mathcal{P}}}
\newcommand{\cle}{{\mathcal{E}}}

\newcommand{\clm}{{\mathcal{M}}}
\newcommand{\xdif}{{\mathrm{d}}}
\newcommand{\rmdd}{{\mathrm{D}}}
\newcommand{\rmc}{{\mathrm{C}}}
\newcommand{\rms}{{\mathrm{S}}}
\newcommand{\psyms}{{\mathsf{\Pi}}_{\mathrm{sym}}^{(s)}}
\newcommand{\psymt}{{\mathsf{\Pi}}_{\mathrm{sym}}^{(t)}}

\newcommand{\ax}{{\mathsf{X}}}
\newcommand{\mm}{{\mathsf{M}}}
\newcommand{\tr}{\mathrm{tr}}
\newcommand{\ron}{{\mathrm{ran}}}
\newcommand{\dig}{\mathrm{diag}}

\newcommand{\zmx}{\mathbf{0}}
\newcommand{\wmu}{\widetilde{\mu}}
\newcommand{\wclb}{\widetilde{\mathcal{B}}}
\newcommand{\wta}{{\tilde{a}}}
\newcommand{\wb}{{\tilde{b}}}

\begin{document}
\clearpage
\preprint{}

\title{Uncertainty and certainty relations for quantum coherence with respect to design-structured POVMs}

\author{Alexey E. Rastegin}
\affiliation{Department of Theoretical Physics, Irkutsk State University, Irkutsk 664003, Russia}

\begin{abstract}
The concept of quantum coherence and its possible use as a resource
are currently the subject of active researches. Uncertainty and
complementarity relations for quantum coherence allow one to study
its changes with respect to other characteristics of the process of
interest. Protocols of quantum information processing often use
measurements that have a special inner structure. Quantum designs
are considered as tools with a perspective of fruitful applications
in quantum information processing. We obtain uncertainty and
certainty relations for coherence averaged with respect to a set of
design-structured POVMs of rank one. To characterize the degree of
coherence, the relative entropy of coherence is utilized. The
derived relations are exemplified with quantum designs in the
two-dimensional Hilbert space.
\end{abstract}


\maketitle

\pagenumbering{arabic}
\setcounter{page}{1}

\section{Introduction}\label{sec1}

The principle of state superposition plays a crucial role in quantum
mechanics. Various aspects of the problem of coherence have found a
considerable attention within historical development of modern
physics. It is now clear that complete understanding of this problem
could be reached only via a purely quantum approach
\cite{bcp14,plenio16}. Efforts to build resource theory of quantum
coherence are a part of more general attempt to resolve the
strengths and limitations of non-classical correlations
\cite{adesso16jpa}. In order to implement efficient and
fault-tolerant quantum computations, an impact of quantum coherence
on manipulations with quantum register should be understood. Results
of recent investigations have supported this conclusion
\cite{hillery16,hfan2016,rastrol,rastflip}.

Complementarity relations for quantum coherence were considered in
\cite{hall15,pzflf16}. Uncertainty relations for quantum coherence
were formulated in several ways \cite{pati16,baieta17,rastcum}.
Since the notion of coherence is basis dependent, complementarity
and uncertainty relations of interest should take into account a
character of utilized measurements. For measurements with a certain
inner structure, we can often improve uncertainty bounds that follow
from the results of general scope. Mutually unbiased bases are an
especially important example of such measurements used for various
purposes \cite{bz10}. Symmetric informationally complete
measurements give another very helpful tool \cite{rbksc04}. Due to
nice properties, quantum $t$-designs have been proposed for
applications in quantum information science
\cite{scottjpa,ambain07}.

The aim of this work is to examine uncertainty and certainty
relations for quantum coherence with respect to POVMs assigned to a
quantum design. The consideration is essentially based on new
entropic uncertainty relations recently derived in \cite{rastpol}.
The paper is organized as follows. In Sec. \ref{sec2}, the
preliminary facts are recalled. The main findings are presented in
Sec. \ref{sec3}. The derived uncertainty and certainty relations are
illustrated with qubit examples. In Sec. \ref{sec4}, we conclude the
paper with a summary of the obtained results.

\section{Definitions and notation}\label{sec2}

In this section, we will present required material concerning the
relative entropy of coherence and some properties of quantum
designs. Let $\hh$ be $d$-dimensional Hilbert space. We denote the
space of linear operators on $\hh$ by $\lnp(\hh)$ and the set of
positive semidefinite operators by $\lsp(\hh)$. By $\ron(\ax)$, we
mean the range of operator $\ax$. The state of a quantum system is
described by the density matrix $\bro\in\lsp(\hh)$ assumed to be
normalized as $\tr(\bro)=1$. Then the von Neumann entropy is written
as
\begin{equation}
\rms_{1}(\bro):=\!{}-\tr(\bro\ln\bro)
\, . \label{vnent}
\end{equation}

A rigorous framework for the quantification of coherence was
developed in \cite{bcp14}. To the given orthonormal basis
$\clb=\bigl\{|b_{j}\rangle\bigr\}$, one assigns the set of all
density matrices that are diagonal in this basis, viz.
\begin{equation}
\bdl=\sum_{j=1}^{d} \delta_{j}\,|b_{j}\rangle\langle{b}_{j}|
\, . \label{incod}
\end{equation}
These density matrices form the set $\cli_{\clb}$ of states
incoherent with respect to $\clb$. To quantify the amount of
coherence with respect to $\clb$, one measures how far the given
state is from states of $\cli_{\clb}$. The authors of \cite{bcp14}
listed general conditions for quantifiers of coherence. Additional
conditions imposed on coherence measures were presented in
\cite{plenio16}. The imposed conditions allow us to select proper
candidates to quantify quantum coherence. In this paper, the
relative entropy of coherence will be utilized.

For density matrices $\bro$ and $\bmg$, the quantum relative entropy
is defined as \cite{wilde17}
\begin{equation}
\rmdd_{1}(\bro||\bmg):=
\begin{cases}
\tr(\bro\ln\bro\,-\bro\ln\bmg) \,,
& \text{if $\ron(\bro)\subseteq\ron(\bmg)$} \, , \\
+\infty\, , & \text{otherwise} \, .
\end{cases}
\label{relan}
\end{equation}
Although the relative entropy cannot be treated as a metric, it is
one of widely used measures of distinguishability of quantum states.
Following \cite{bcp14}, one defines the coherence measure
\begin{equation}
\rmc_{1}(\clb|\bro):=
\underset{\bdl\in\cli_{\clb}}{\min}\,\rmdd_{1}(\bro||\bdl)
\, . \label{c1df}
\end{equation}
The minimization procedure results in the formula \cite{bcp14}
\begin{equation}
\rmc_{1}(\clb|\bro)=\rms_{1}(\bro_{\dig})-\rms_{1}(\bro)
\, , \label{c1for}
\end{equation}
where the diagonal state
\begin{equation}
\bro_{\dig}:=\dig\bigl(
\langle{b}_{1}|\bro|b_{1}\rangle,\ldots,\langle{b}_{d}|\bro|b_{d}\rangle
\bigr)
\, . \label{rhodig}
\end{equation}
We can represent $\rms_{1}(\bro_{\dig})$ as the Shannon entropy
calculated with probabilities
$p_{j}(\clb|\bro)=\langle{b}_{j}|\bro|b_{j}\rangle$:
\begin{equation}
\rms_{1}(\bro_{\dig})=H_{1}(\clb|\bro):=-\sum_{j=1}^{d} p_{j}(\clb|\bro)\, \ln{p}_{j}(\clb|\bro)
\, . \label{shlin}
\end{equation}
For basic properties of (\ref{c1df}), see the relevant sections of
\cite{bcp14,plenio16}. The relative entropy of coherence seems to be
the most justifiable measure. Together with (\ref{relan}), other
quantum divergences were considered, including the quasi-entropies
of Petz \cite{petz86}. It is for this reason that we designate the
considered entropic quantities by the subscript $1$. Coherence
quantifiers induced by quantum divergences of the Tsallis type were
addressed in \cite{rastalp}. It turned out that such quantifiers do
not have a simple form similar to (\ref{c1for}). Coherence monotones
based on R\'{e}nyi divergences were considered in
\cite{chitam2016,shao16,skwgb16}.

The above definition is related to the case of orthonormal bases. In
quantum information science, generalized quantum measurements are an
indispensable tool. Let $\clm=\bigl\{\mm_{j}\bigr\}$ be a set of
elements $\mm_{j}\in\lsp(\hh)$ satisfying the completeness relation
\begin{equation}
\sum\nolimits_{j}   \mm_{j}=\pen_{d}
\, , \label{conren}
\end{equation}
where $\pen_{d}$ is the identity operator on $\hh$. These elements
form a positive operator-valued measure (POVM) \cite{wilde17}. It is
important that the number of possible outcomes can exceed the
dimensionality. The question of quantifying quantum coherence beyond
the case of orthonormal bases was addressed in
\cite{rastlud,bkb2019,dsende}. In the following, we will deal only
with rank-one POVMs, which play very important role in quantum
information theory. One of the reasons for their utility was
revealed by Davies \cite{davies78}. Let
$\bigl\{|\mu_{j}\rangle\bigl\}_{j=1}^{N}$ be a set of sub-normalized
vectors that form a rank-one POVM with elements
$\mm_{j}=|\mu_{j}\rangle\langle\mu_{j}|$. By $\mu_{ij}$, we mean
$i$-th component of $j$-th vector $|\mu_{j}\rangle$ in the
calculation basis. Due to (\ref{conren}), rows of the
$d\times{N}$-matrix $[[\mu_{ij}]]$ are mutually orthogonal. By
adding $(N-d)$ new rows, this matrix can be converted into a unitary
$N\times{N}$-matrix. Its columns denoted by $|\wmu_{j}\rangle$ form
an orthonormal basis $\wclb$ in $N$ dimensions. As a block matrix,
each column is now written as
\begin{equation}
|\wmu_{j}\rangle:=
\begin{pmatrix}
 |\mu_{j}\rangle \\
 |\mu_{j}^{\prime}\rangle
\end{pmatrix}
 . \label{wum}
\end{equation}
To the original density matrix, we assign $\wbro=\dig(\bro,\zmx)$,
so that
$\langle\wmu_{j}|\wbro|\wmu_{j}\rangle=\langle\mu_{j}|\bro|\mu_{j}\rangle$
for $1\leq{j}\leq{d}$. Following \cite{rastlud}, we characterize the
amount of coherence of $\bro$ with respect to
$\bigl\{|\mu_{j}\rangle\bigl\}_{j=1}^{N}$ by the quantity
\begin{equation}
\rmc_{1}(\clm|\bro):=\rmc_{1}(\wclb|\wbro)=
H_{1}(\wclb|\wbro)-\rms_{1}(\wbro)=H_{1}(\clm|\bro)-\rms_{1}(\bro)
\, . \label{c1wclb}
\end{equation}
Thus, the relative entropy of coherence is expressed in terms
related to the original space $\hh$ solely. The authors of
\cite{bkb2019} showed that the relative entropy of coherence does
not depend on the choice of a Naimark extension for its definition.
The right-hand side of (\ref{c1wclb}) should be modified, when we do
not restrict a consideration to rank-one POVMs.

Let us recall some facts about quantum designs. In 
$d$-dimensional Hilbert space $\hh$, we consider lines passing
through the origin. These lines form a complex projective space
\cite{scottjpa}. Up to a phase factor, each line is represented by a
unit vector $|\phi\rangle\in\hh$. The set
$\dn=\bigl\{|\phi_{k}\rangle:\,\langle\phi_{k}|\phi_{k}\rangle=1,\,k=1,\ldots,K\bigr\}$
is a quantum $t$-design, when for all real polynomials $\calp_{t}$
of degree at most $t$ it holds that
\begin{equation}
\frac{1}{K^{2}}\sum_{j,k=1}^{K}
\calp_{t}\Bigl(\bigl|\langle\phi_{j}|\phi_{k}\rangle\bigr|^{2}\Bigr)=
\int\!\int\xdif\mu(\psi)\,\xdif\mu(\psi^{\prime})\>
\calp_{t}\Bigl(\bigl|\langle\psi|\psi^{\prime}\rangle\bigr|^{2}\Bigr)
\, . \label{tdesdf}
\end{equation}
Here, $\mu(\psi)$ denotes the unique unitarily-invariant probability
measure on the corresponding complex projective space
\cite{scottjpa}. It follows from the definition that each $t$-design
is also a $s$-design with $s\leq{t}$. In general, $t$-designs in
projective spaces were examined in \cite{hoggar82}. Due to findings
of the paper \cite{seym1984}, we know that quantum $t$-designs exist
for all suitable $t$ and $d$. However, these results do not provide
a common strategy to generate designs in all respective cases. In
effect, there are several explicit examples to test a theoretical
framework.

Quantum designs have interesting formal properties. Let $\psymt$ be
the projector onto the symmetric subspace of $\hh^{\otimes{t}}$. The
trace of $\psymt$ gives dimensionality of this symmetric subspace. It
holds that \cite{scottjpa}
\begin{equation}
\frac{1}{K}\>\sum_{k=1}^{K}
|\phi_{k}\rangle\langle\phi_{k}|^{\otimes{t}}=\cald_{d}^{(t)}\,\psymt
\, , \label{topys}
\end{equation}
where $\cald_{d}^{(t)}$ denotes the inverse of $\tr\bigl(\psymt\bigr)$, namely
\begin{equation}
\cald_{d}^{(t)}=\binom{d+t-1}{t}^{\!-1}
=\frac{t!\,(d-1)!}{(d+t-1)!}
\ . \label{indim}
\end{equation}
At the given $t$, we can rewrite (\ref{topys}) for all positive
integers $s\leq{t}$. Substituting $t=1$ leads to the formula
\begin{equation}
\frac{d}{K}\>\sum_{k=1}^{K}
|\phi_{k}\rangle\langle\phi_{k}|=\pen_{d}
\, . \label{comprel}
\end{equation}
Thus, unit vectors $|\phi_{k}\rangle$ allow us to build to a
resolution of the identity in $\hh$. In principle, there may be
several resolutions assigned to the given $t$-design. We cannot list
all of them {\it a priori}, without an explicit analysis of
$|\phi_{k}\rangle$. Obviously, one can take the complete set $\cle$
consisting of operators
\begin{equation}
\me_{k}=\frac{d}{K}\>
|\phi_{k}\rangle\langle\phi_{k}|
\, . \label{mkdef}
\end{equation}
Sometimes, $M$ rank-one POVMs $\bigl\{\cle^{(m)}\bigr\}_{m=1}^{M}$
can be assigned to the given quantum design. Each of them
consist of $\ell$ operators of the form
\begin{equation}
\me_{j}^{(m)}=\frac{d}{\ell}\>
|\phi_{j}^{(m)}\rangle\langle\phi_{j}^{(m)}|
\, . \label{mejdf}
\end{equation}
The integers $\ell$ and $M$ are connected by $K=\ell{M}$.

If the state of interest is described by density matrix $\bro$, then
the probability of $j$-th outcome is equal to
\begin{equation}
p_{j}(\cle^{(m)}|\bro)=\frac{d}{\ell}\,\langle\phi_{j}^{(m)}|\bro|\phi_{j}^{(m)}\rangle
\, . \label{probk}
\end{equation}
It follows from (\ref{topys}) that for all $s=2,\ldots,t$ we have
\cite{guhne19}
\begin{equation}
\frac{1}{K}\>\sum_{k=1}^{K}
\langle\phi_{k}|\bro|\phi_{k}\rangle^{s}=\cald_{d}^{(s)}\tr\bigl(\bro^{\otimes{s}}\psyms\bigr)
\, . \label{indet}
\end{equation}
Combining (\ref{probk}) with (\ref{indet}) then gives
\begin{equation}
\sum_{m=1}^{M}\sum_{j=1}^{\ell}p_{j}(\cle^{(m)}|\bro)^{s}=
\left(\frac{d}{\ell}\right)^{\!s}\,\sum_{k=1}^{K}\langle\phi_{k}|\bro|\phi_{k}\rangle^{s}=
K\ell^{-s}d^{\,s}\,\cald_{d}^{(s)}\tr\bigl(\bro^{\otimes{s}}\psyms\bigr)
\, . \label{mindet}
\end{equation}
When a single POVM is assigned, one has $\ell=K$ and
\begin{equation}
\sum_{k=1}^{K}p_{k}(\cle|\bro)^{s}=
K^{1-s}d^{\,s}\,\cald_{d}^{(s)}\tr\bigl(\bro^{\otimes{s}}\psyms\bigr)
\, . \label{indek}
\end{equation}
The authors of \cite{guhne19,cirac18} have answered the question how
to express $\tr\bigl(\bro^{\otimes{s}}\psyms\bigr)$ as a sum of
monomials of the moments of $\bro$. Of course, complexity of such
expressions increases with growth of $s$. To avoid bulky expressions
in the following, we introduce the two quantities
\begin{align}
\bar{\beta}_{\ell}^{(s)}(\bro)&=\ell^{1-s}d^{\,s}\,\cald_{d}^{(s)}\tr\bigl(\bro^{\otimes{s}}\psyms\bigr)
\, , \label{betn}\\
\bar{\beta}^{(s)}(\bro)&=K^{1-s}d^{\,s}\,\cald_{d}^{(s)}\tr\bigl(\bro^{\otimes{s}}\psyms\bigr)
\, . \label{betk}
\end{align}
The latter is obtained from (\ref{betn}) by taking $\ell=K$. The
formulas (\ref{mindet}) and (\ref{indek}) impose restrictions on
measurement statistics for design-structured POVMs. Uncertainty
relations in terms of generalized entropies were considered in
\cite{guhne19,rastdes}. The paper \cite{rastdes} actually develops
the idea originally proposed in \cite{rastmubs}. Uncertainty and
certainty relations in terms of the corresponding Shannon entropies
are derived in \cite{rastpol}.

\section{Uncertainty and certainty relations for quantum coherence}\label{sec3}

In this section, we will derive uncertainty and certainty relations
for the relative entropy of coherence averaged with respect to a set
of design-structured POVMs. The consideration is essentially based
on the results of \cite{rastpol}. The principal idea is to use
effectively restrictions of the form (\ref{mindet}). Before
presenting the desired complementarity relations, some definitions
from \cite{rastpol} will be recalled. First, we introduce two
families of coefficients, namely
\begin{align}
&a_{n}^{(1)}=\sum_{r=1}^{n-1}\frac{1}{r}
\ , \qquad
&a_{n}^{(s)}=(-1)^{s-1}\sum_{r=s-1}^{n-1}\frac{1}{r}\>\binom{r}{s-1}
\qquad
(2\leq{s}\leq{n})
\, , \label{aeff}\\
&b_{n}^{(1)}=\sum_{r=2}^{n-1}\frac{1}{r}
\, , \qquad
&b_{n}^{(s)}=\frac{(-1)^{s-1}}{s}\sum_{r=s-1}^{n-1}\frac{1}{r}\>\binom{r}{s-1}
\qquad
(2\leq{s}\leq{n})
\, . \label{boeff}
\end{align}
These coefficients follow from truncated expansions according to the
Taylor formula \cite{rastpol}. Another two families of coefficients
are represented in terms of coefficients of $n$-th shifted Chebyshev
polynomials
\begin{equation}
c_{n}^{(s)}=(-1)^{n+s}\,2^{2s-1}
\!\left[
2\binom{n+s}{n-s}-\binom{n+s-1}{n-s}
\right]
 . \label{ccoeff}
\end{equation}
They are used to obtain expansions with flexible coefficients. Such
expansions were introduced and motivated by Lanczos \cite{lanczos}.
Further, we define the coefficients \cite{rastpol}
\begin{align}
&\wta_{n}^{(1)}=
\frac{(-1)^{n}}{2n^{2}}\,\sum_{s=2}^{n} \frac{c_{n}^{(s)}}{s-1}
\ , \qquad
\wta_{n}^{(s)}=\frac{(-1)^{n+1}}{2n^{2}}\,\frac{c_{n}^{(s)}}{s-1}
\qquad
(2\leq{s}\leq{n})
\, . \label{waeff}\\
&\wb_{n}^{(0)}=1-\sum_{s=1}^{n} \frac{\wta_{n}^{(s)}}{s}
\ , \qquad
\wb_{n}^{(1)}=\wta_{n}^{(1)}-1
\, , \qquad
\wb_{n}^{(s)}=\frac{\wta_{n}^{(s)}}{s}
\qquad
(2\leq{s}\leq{n})
\, . \label{wbeff}
\end{align}
The following result will be proved in \cite{rastdes}. Let
$\Upsilon_{K-1}^{(t)}(\beta)$ denote the maximal real root of the
algebraic equation
\begin{equation}
(1-\Upsilon)^{t}+(K-1)^{t-1}\Upsilon^{t}=(K-1)^{t-1}\beta
\, . \label{curve}
\end{equation}
For all $k=1,\ldots,K$ we have \cite{rastdes}
\begin{equation}
p_{k}(\cle|\bro)\leq\Upsilon_{K-1}^{(t)}\bigl(\bar{\beta}^{(t)}(\bro)\bigr)
\, . \label{pkabov}
\end{equation}
Complementarity relations for the relative entropy of coherence with
respect to design-structured POVMs are posed as follows.

\newtheorem{prop1}{Proposition}
\begin{prop1}\label{rst1}
Let $M$ rank-one POVMs $\cle^{(m)}$, each with $\ell$ elements of
the form (\ref{mejdf}), be assigned to a quantum $t$-design
$\dn=\bigl\{|\phi_{k}\rangle\bigr\}_{k=1}^{K}$ in $d$ dimensions. It
then holds that
\begin{align}
\sum_{s=1}^{t} a_{t}^{(s)}\Upsilon^{1-s}\bar{\beta}_{\ell}^{(s)}(\bro)
-\ln\Upsilon-\rms_{1}(\bro)
&\leq\frac{1}{M}\sum_{m=1}^{M}\rmc_{1}(\cle^{(m)}|\bro)
\leq\frac{\Upsilon\ell}{t}+
\sum_{s=1}^{t} b_{t}^{(s)}\Upsilon^{1-s}\bar{\beta}_{\ell}^{(s)}(\bro)-
\ln\Upsilon-\rms_{1}(\bro)
\, , \label{cotay}\\
\sum_{s=1}^{n}\wta_{n}^{(s)}\Upsilon^{1-s}\bar{\beta}_{\ell}^{(s)}(\bro)
-\ln\Upsilon-\rms_{1}(\bro)
&\leq\frac{1}{M}\sum_{m=1}^{M}\rmc_{1}(\cle^{(m)}|\bro)
\leq\wb_{n}^{(0)}\Upsilon\ell+
\sum_{s=1}^{n}\wb_{n}^{(s)}\Upsilon^{1-s}\bar{\beta}_{\ell}^{(s)}(\bro)
-\ln\Upsilon-\rms_{1}(\bro)
\, . \label{coche}
\end{align}
Here
$\Upsilon=\min\bigl\{M\,\Upsilon_{K-1}^{(t)}\bigl(\bar{\beta}^{(t)}(\bro)\bigr),1\bigr\}$,
$n=\min\{t,15\}$ and the coefficients are respectively defined by
(\ref{aeff})--(\ref{wbeff}).
\end{prop1}

{\bf Proof.} Under the same preconditions, it was proved in \cite{rastpol} that
\begin{align}
\sum_{s=1}^{t} a_{t}^{(s)}\Upsilon^{1-s}\bar{\beta}_{\ell}^{(s)}(\bro)
-\ln\Upsilon
&\leq\frac{1}{M}\sum_{m=1}^{M}H_{1}(\cle^{(m)}|\bro)\leq\frac{\Upsilon\ell}{t}+
\sum_{s=1}^{t} b_{t}^{(s)}\Upsilon^{1-s}\bar{\beta}_{\ell}^{(s)}(\bro)-
\ln\Upsilon
\, , \label{twosid}\\
\sum_{s=1}^{n}\wta_{n}^{(s)}\Upsilon^{1-s}\bar{\beta}_{\ell}^{(s)}(\bro)
-\ln\Upsilon
&\leq\frac{1}{M}\sum_{m=1}^{M}H_{1}(\cle^{(m)}|\bro)
\leq\wb_{n}^{(0)}\Upsilon\ell+
\sum_{s=1}^{n}\wb_{n}^{(s)}\Upsilon^{1-s}\bar{\beta}_{\ell}^{(s)}(\bro)
-\ln\Upsilon
\, . \label{tosid}
\end{align}
Combining (\ref{c1wclb}) with (\ref{twosid}) and (\ref{tosid}) leads
to (\ref{cotay}) and (\ref{coche}), respectively. $\blacksquare$

The statement of Proposition \ref{rst1} provides two-sided estimates
on the averaged relative entropy of coherence taken with respect to
a set of design-structured POVMs. It was found in \cite{rastpol}
that the two-sided estimates (\ref{twosid}) and (\ref{tosid}) give
better results in application to single assigned POVM. In the case
of rank-one POVM $\cle$ with $K$ elements of the form (\ref{mkdef}),
the results (\ref{cotay}) and (\ref{coche}) read as
\begin{align}
\sum_{s=1}^{t}a_{t}^{(s)}\Upsilon^{1-s}\bar{\beta}^{(s)}(\bro)
-\ln\Upsilon-\rms_{1}(\bro)
&\leq\rmc_{1}(\cle|\bro)
\leq\frac{\Upsilon{K}}{t}+
\sum_{s=1}^{t} b_{t}^{(s)}\Upsilon^{1-s}\bar{\beta}^{(s)}(\bro)-
\ln\Upsilon-\rms_{1}(\bro)
\, , \label{cotayk}\\
\sum_{s=1}^{n}\wta_{n}^{(s)}\Upsilon^{1-s}\bar{\beta}^{(s)}(\bro)
-\ln\Upsilon-\rms_{1}(\bro)
&\leq\rmc_{1}(\cle|\bro)
\leq\wb_{n}^{(0)}\Upsilon{K}+
\sum_{s=1}^{n}\wb_{n}^{(s)}\Upsilon^{1-s}\bar{\beta}_{\ell}^{(s)}(\bro)
-\ln\Upsilon-\rms_{1}(\bro)
\, , \label{cochek}
\end{align}
where
$\Upsilon=\Upsilon_{K-1}^{(t)}\bigl(\bar{\beta}^{(t)}(\bro)\bigr)$.
For the maximally mixed state $\bro_{*}=\pen_{d}/d$, the above
bounds are saturated. In very deed, the two-sided estimates
(\ref{twosid}) and (\ref{tosid}) are saturated here \cite{rastpol}.

\begin{figure*}
\begin{center}
\includegraphics[height=7.6cm]{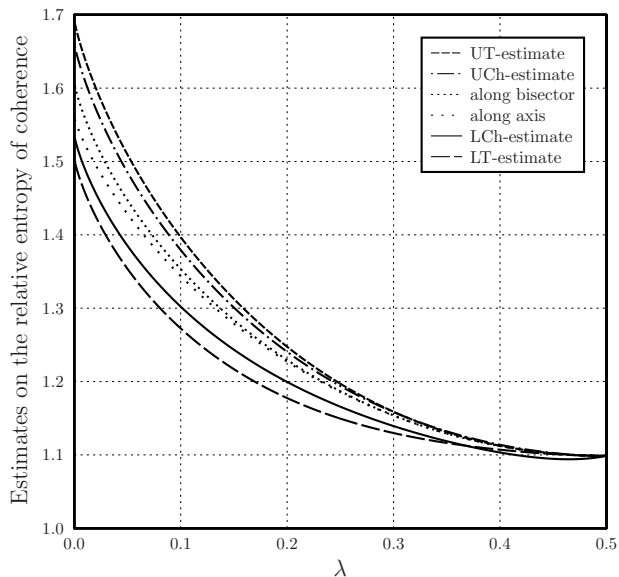}
\caption{\label{fig1} Estimates on the relative entropy of coherence versus $\lambda$ for the $3$-design with $6$ vertices.}
\end{center}
\end{figure*}

We shall apply the above complementarity relations to concrete
quantum designs in two dimensions. The authors of \cite{guhne19}
gave a short description of these designs in terms of components of
the Bloch vector. To represent each $|\phi_{k}\rangle$, the Bloch
vector comes to one of vertices of some polyhedron. The qubit
density matrix is characterized by its minimal eigenvalue $\lambda$.
We also restrict a consideration to the case of single assigned
POVM. To avoid bulky legends on figures, the following notation will
be used. By ``LT-estimate'' and ``UT-estimate'', we mean the left-
and right-hand sides of (\ref{cotayk}), respectively. They are based
on approximation by polynomials, whose coefficients are due to the
Taylor scheme. The terms ``LCh-estimate'' and ``UCh-estimate''
respectively refer to the left- and right-hand sides of
(\ref{cochek}). These estimates use polynomials with coefficients
linked to coefficients of the shifted Chebyshev polynomials.

\begin{figure*}
\begin{center}
\includegraphics[height=7.6cm]{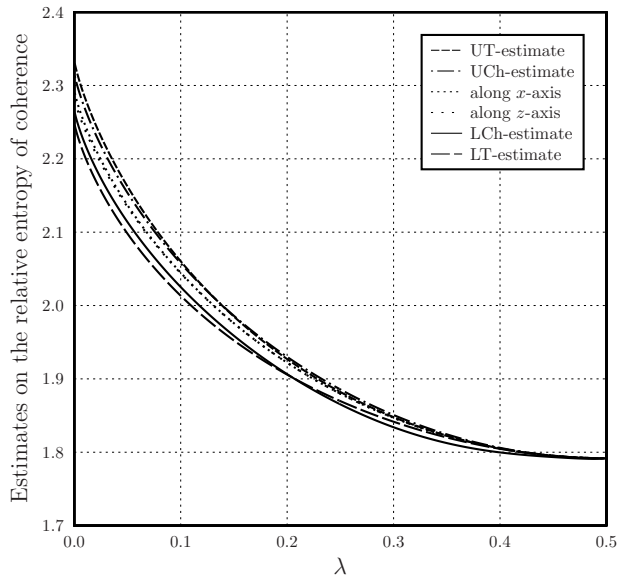}
\end{center}
\caption{\label{fig2} Estimates on the relative entropy of coherence versus $\lambda$ for the $5$-design with $12$ vertices.}
\end{figure*}

First, we consider estimates on coherence with respect to POVM
assigned to the $3$-design with $K=6$ vertices of octahedron. In
Fig. \ref{fig1}, we plot the two lower estimates and the two upper
ones as functions of $\lambda$. The relative entropy of coherence
changes in sufficiently restricted diapason. By two dotted lines, we
also show values of the relative entropy of coherence for two
orientations of the Bloch vector. Let the three coordinate axes pass
through vertices of octahedron. One of the used orientations is
along a coordinate axis, whereas the other lies in a coordinate
plane along quadrant bisector. The two-sided estimate (\ref{cotayk})
is better for states sufficiently close to the maximally mixed one.
The formula (\ref{cochek}) gives a stronger bound for pure states
and states with moderate mixedness. Near the right least point
$\lambda=1/2$, all the curves converge at one point.

\begin{figure*}
\begin{center}
\includegraphics[height=7.6cm]{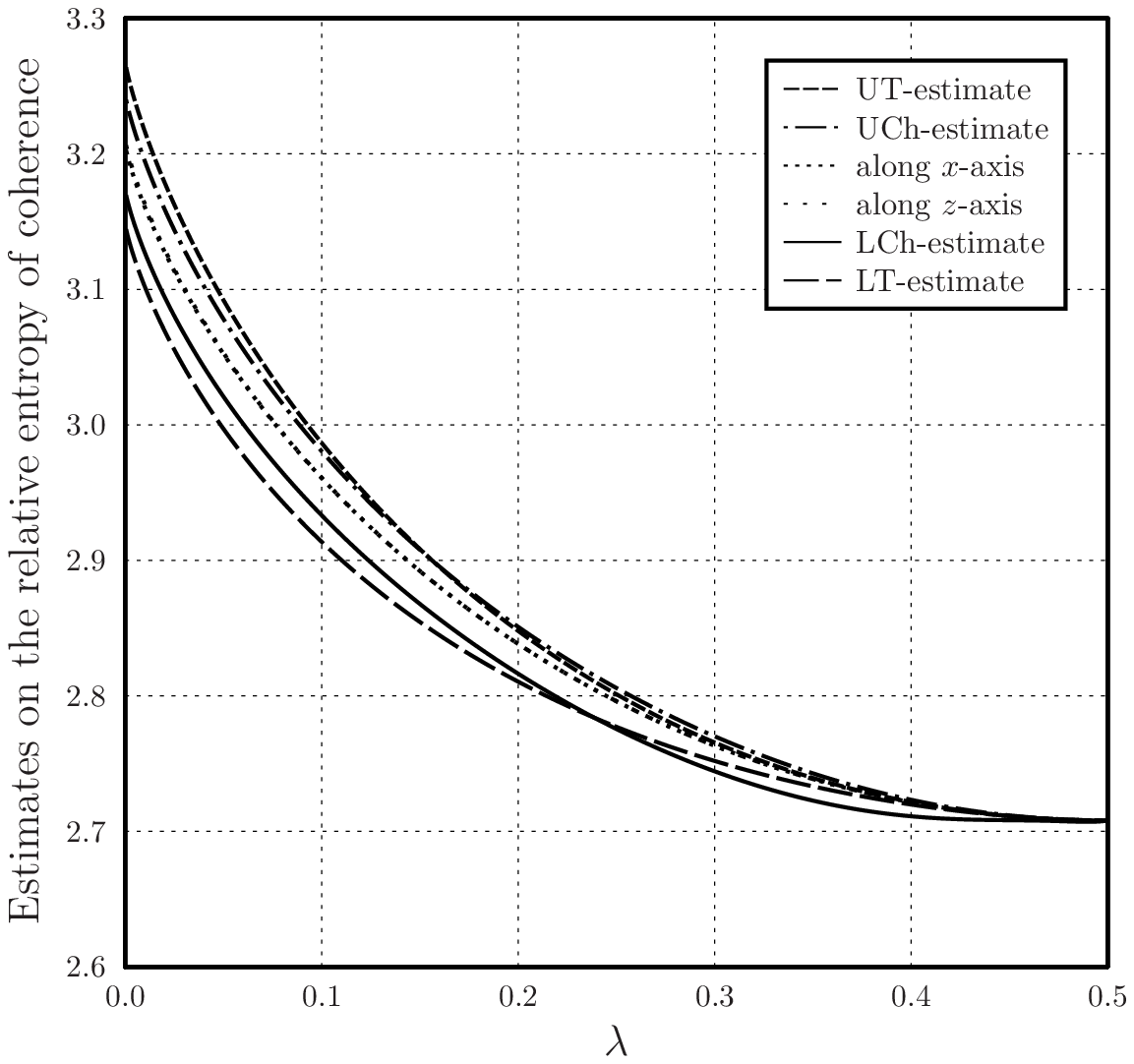}
\end{center}
\caption{\label{fig3} Estimates on the relative entropy of coherence versus $\lambda$ for the $5$-design with $30$ vertices.}
\end{figure*}

The following example concerns the $5$-design with $K=12$ vertices
forming an icosahedron. The four estimates on the relative entropy
of coherence versus $\lambda$ are shown in Fig. \ref{fig2}. Of
course, possible changes of the relative entropy of coherence lie in
enough limited diapason. Similarly to the previous example, the
two-sided estimate (\ref{cotayk}) is insufficient for pure states
and states with low mixedness. One the other hand, we see wider
region in which the result (\ref{cotayk}) leads to stronger bounds.
By two dotted lines, we also show values of the relative entropy of
coherence for two orientations of the Bloch vector. Let the $z$-axis
pass through two opposite vertices and form symmetry axis of
icosahedron. Let the $x$-axis pass so that one of inclined edges
lies in the $zx$-plane. One of the used orientations is along the
$z$-axis, whereas the other is along positive part of the $x$-axis.
In this example, the two dotted lines are closer than in Fig.
\ref{fig1}. Again, all the curves converge at one point for
$\lambda=1/2$.

Finally, we address the $5$-design with $K=30$ vertices forming an
icosidodecahedron. In Fig. \ref{fig3}, the two lower estimates and
the two upper ones are shown as functions of $\lambda$. Due to
increasing number of vertices, the relative entropy of coherence
generally takes larger values than in the two previous examples.
Nevertheless, main observations remain the same. By two dotted
lines, one shows values for two orientations of the Bloch vector.
Let the $z$-axis pass through two opposite pentagons, and let the
$x$-axis pass through one of equatorial vertices. One of the used
orientations is along the $z$-axis, whereas the other is along
positive part of the $x$-axis. Here, the dotted lines are almost
indistinguishable. It witnesses that  some improvements of
uncertainty and certainty bounds may be therein. However, such
improvements would give a correction in few percents. This question
deserves further investigations.

With growth of $t$, we have seen narrowing of the range in which the
relative entropy of coherence varies. It is a reflection of
increasing number of imposed restrictions. Applying the two-sided
estimates (\ref{cotayk}) and (\ref{cochek}) to other examples of
quantum designs in two dimensions support the above observations.
For states with moderate mixedness, the formula (\ref{cochek})
should be preferred. Recall that it is based on power expansions
with flexible coefficients. For sufficiently mixed states, the
two-sided estimate (\ref{cotayk}) provides better results. This
estimate follows from truncated expansions according to the Taylor
formula. The interval $\lambda\in[0,1/2]$ is divided into two parts
unequal generally. The former is where the two-sided estimate
(\ref{cochek}) is stronger, and the latter is a domain for the use
of (\ref{cotayk}). The second part becomes wider when $t$ grows.

\section{Conclusions}\label{sec4}

We have considered uncertainty bounds on the relative entropy of
coherence averaged with respect to a set of design-structured POVMs.
The concept of coherence in purely quantum formulation is currently
the subject of active researches. It is also well known that
generalized quantum measurements are an indispensable tool in
quantum information science. One of principal properties of such
measurements is that the number of possible outcomes can exceed the
system dimensionality. Among generalized measurements, rank-one
POVMs form an especially important class. A quantum design consists
of unit vectors that enjoy a list of formal properties. In
particular, rank-one POVMs can be built of these vectors. Hence, the
question of characterizing coherence quantifiers with respect to
design-structured POVMs is interesting for several reasons. The
derived relations are mainly based on novel uncertainty and
certainty relations derived recently in \cite{rastpol}. The inner
structure of assigned POVMs leads to sufficiently strong
restrictions on measurement statistics. On the other hand, such
restrictions allow one to estimate the relative entropy of coherence
from below as well as from above. As a result, the relative entropy
of coherence with respect to design-structured POVMs varies in
enough narrow diapason. The considered examples also allow one to
compare two ways to construct two-sided estimates on the entropic
functions of interest.

\end{document}